\documentstyle[11pt,newpasp,twoside,epsf]{article}
\def\edcomment#1{\iffalse\marginpar{\raggedright\sl#1\/}\else\relax\fi}
\reversemarginpar

\begin{document}

\null
\null

\noindent Proceedings of an International Conference on Scientific Frontiers in Research on Extrasolar PlanetsSymposium, held on June 18-21, 2002, in
Washington DC, USA, ASP Conference Series, vol. 294, 2003, Editor Drake Deming and Seager, p. 587-590. 

\null
\null

\title{Future Astrometry with ALMA to characterise extra-solar planet 
orbits}
  \author{Jean-Fran\c cois Lestrade}
\affil{Observatoire de Paris - CNRS, 77 av. Denfert Rochereau,
F75014 - Paris - France, Jean-francois.lestrade@obspm.fr}

\begin{abstract}
The most complete catalogue of nearby stars by Gliese and Jarheiss (1991)
is used to find that the thermal emission of the photospheres of 
446 nearby stars can be detected  by the future millimeter array ALMA. 
This array has a theoretical astrometric precision of 0.1 milliarcsond. 
A long-term astrometric observation programme  ($\ge$ 10 yr)
of these stars would be 
sensitive to wobbles caused by the gravitational pull of possible unseen 
planets. Such a programme would  probe   
 minimum planetary masses as low as 0.1 Jupiter for an orbital 
period of 10 years. 
We provide the histogramme of these minimum masses for the 446 ALMA stars.   
\end{abstract}

\section{Introduction}

Discovery of Jupiter-mass planets around other stars 
 by the radial velocity technique (Doppler) since 1995 has been one 
of the  highlights of the last decade in astronomy.   
Astrometry, similarly to the radial velocity technique, is 
an indirect method that  detects 
the reflex motion (wobble) of the central star caused by the gravitational
pull of the orbiting planet(s). 
Even when direct imaging of extra-solar planets is routine  in the future,
high-precision Doppler and astrometry will still  
be used  to measure key parameters,
{\sl e.g.} masses, mutual inclinations and node of orbits
to characterise  planetary systems. These last two
parameters yield importantly the degree of coplanarity 
of orbits in multiple planet systems.

ALMA is an interferometric
array of 64 antennas operating at radio millimeter wavelengths that is
planned for construction at Chajnantor (altitude = 5000 m) in Chili.
This ground-based facility will be completed in 2010. 

We searched the stars that can be detected by ALMA at millimeter wavelengths
in the most complete catalogue  
of Nearby Stars (CNS3 3rd Edition) by Gliese 
and Jarheiss (1991). We assess the astrometric potential of ALMA for indirect
detection of extra-solar planets possibly orbiting these stars.

\section{ The catalogue CNS3 of Nearby Stars   }

There are  3461 stars in the  CNS3 that are mostly at $\le 25$~pc and have   
spectral type and magnitude V measured. There are 341 additional
stars  lacking either spectral type or magnitude V. 
The CNS3 is more complete than Hipparcos for nearby stars but is thought still 
 uncomplete  even within  10 parsecs. 

\begin{center}
\begin{tabular}{l|l|l|l|l} \hline
Spectral    & Nber of stars       & Nber of stars &  Nber of stars    & Fraction  \\ 
Type        & in the whole CNS3   & in the CNS3   & detectable        & of the    \\
            & (mostly d $\le 25$ pc) & at d $\le 10 $ pc    & by ALMA &  CNS3     \\  \hline
      O     &   ~~~0              &     ~~0       & ~~0     & ~0  \%              \\
      B     &   ~~~3              &     ~~1       & ~~2     & 66  \%              \\
      A     &   ~~69              &     ~~5       & ~54     & 78  \%              \\
      F     &   ~266              &     ~11       & 158     & 59  \%              \\
      G     &   ~495              &     ~30       & 125     & 35  \%              \\
      K     &   ~824              &     ~57       & ~71     & ~9  \%              \\
      M     &    1804             &     291       & ~36     & ~2  \%              \\  \hline
      Total &    3461             &     395       & 446     & 13  \%              \\ \hline
\end{tabular}
\end{center}

\null

\noindent {\sl  Table 1 : Distribution of spectral type in the catalogue CNS3.
Distribution of stars detectable by ALMA above 0.1 milliJansky at 345~GHz.
For some stars, the catalogue  provides only  
 approximate spectral types that have been interpretated
in our analysis as the following~:
 a-f=F0, f=F5, f-g=G0, g=G5, g-k=K0, k=K5, k-m=M0, m=M2, m+=M6.}

\section{Thermal emission of stars detectable by ALMA and astrometric precision }

ALMA can detect the thermal emission from the photospheres 
of some stars at radio millimeter wavelengths. This  provides a well-defined 
surface whose radio centroid position can accurately track the reflex motion of 
an unseen companion. Is a spotted photosphere a limiting factor ?. 
If star spots were  present there would be  an 
additional position modulation with the known rotation period of the star.  
If a G~dwarf  were covered by spots over 0.2 \% of its surface
and the differential temperature photosphere$-$spot be $\Delta$~T=1000~K,
typical of the Sun, the astrometric modulation amplitude $\xi$ 
would be $\sim~0.1~\mu$arcsec at 10~pc, {\it i.e.} negligeable.
For an M~dwarf covered over 70 \% and $\Delta$~T=1000~K, 
the amplitude  $\xi$ would be $\sim~35~\mu$arcsec at 10~pc.

We have computed the thermal flux densities at 345~GHz for the photospheres  
of all the stars of the CNS3.
The frequency 345~GHz is the optimum sensitivity of ALMA for this project.
 This calculation 
is based on the Planck radiation law; the brightness is  
$ {\rm B}(\nu,T) = {{2 h \nu^3} \over {c^2}} 
{ 1 \over  { e^{h \nu / kT } -1 } } $; 
 the unit of $ {\rm B}(\nu,T)$ is W~m$^{-2}$~Hz$^{-1}$~Rd$^{-2}$  
with the proper constants. The  spectral types  
from the catalogue were used to derive the effective temperatures $T$ 
and the linear diameters of the stars.
 The distance was taken also  from the catalogue to compute
 the photospheric surface  $\Omega$ in Rd$^2$ and 
the flux density $ {\rm F}_\nu = {\rm B}(\nu,T) \times \Omega$.

The sensitivity of ALMA yields a signal-to-noise SNR of 30 for
the observation of a solar type star (T$_{eff}$=6500 K and diameter 
= 2R$_{\odot}$) at 10 pc in one-hour of integration at 345 GHz 
($\lambda = 0.87$ mm).

The theoretical astrometric precision of ALMA is :

$$\sigma_{\alpha, \delta} = {1 \over {2 \pi}} ~ {1 \over SNR } 
 ~ { \lambda \over B}  =  0.1~{\rm  milliarcsecond} $$ 

\noindent for  baseline length $B=10$~km,
$\lambda = 0. 87$~mm  and SNR=30. This is 10 times better than 
the astrometric precision of Hipparcos.

Proper calibration scheme should be developped to monitor the
atmospheric phase fluctuation in order to reach this theoretical 
precision. This is a challenging issue.

\section{RESULTS }

\subsection {446 nearby stars  observabe astrometricly by ALMA   }

We have found that 446 nearby stars of the CNS3 have a flux density 
$\ge$~0.1~mJy at 345~GHz and, hence, are detectable by ALMA.
The signal-to-noise ratio SNR 30, required for the astrometric 
precision of 0.1 milliarcsecond,  can be reached 
with an integration time of  a few hours  for  these stars 
by using the full bandwidth (16 GHz) of the array and both polarizations.
The declination cutoff  $\delta < + 35^{\circ}$ was also used in this
study because  ALMA is located at Chajnantor in Chili. The last 2 columns in 
Table 1 show the spectral
types of these ALMA stars and their fractions of the whole CNS3.  The histogramme
of all the integration times (for an SNR=30) of this sample of stars is in Figure~1.




\subsection {Minimum masses for unseen planets detectable astrometricly by ALMA  }

The amplitude of the wobble of the central star due to the gravitational
pull of the planet in a circular orbit is :

$$  \theta =  {m \over M }  { a \over d } = {m \over  d } \Bigg({ P \over M }\Bigg)^{2/3} $$

\noindent  where $\theta$ is in arcsecond, $a$ in AU, $P$ in years, $d$ in pc,
$m$ and $M$, the masses of the unseen companion and of the central star,
in M$_{\odot}$.

We have computed the mass $m$ of the  companion with the
following assumptions :

\noindent 1) astrometric precision of the individual ALMA observation $\sigma = 0.1$
 milliarcsecond.

\noindent 2) the wobble $\theta$  detectable by several observations 
over  at least an orbital period is $\theta = 1 \sigma$. 

\noindent 3) Obital period $P$ = 10 years

\noindent 4) Observation program at least as long as $P$ = 10 years.

\null

The histogramme below summarises the range of minimum masses that can be
probed by ALMA.

\section{ Conclusion}

We found  that 446 nearby
stars of the  Gliese \& Jarheiss catalogue
 can be observed astrometricly by ALMA at radio millimeter wavelengths. 
The theoretical astrometric precision  of  0.1 milliarcsecond 
 can be reached in a  reasonable integration 
time ($\le$ a few hours) for these stars. The challenge to reach this
precision 
is the  monitoring of the rapid phase fluctuations  caused by the
atmosphere even at an altitude of 5000~m.
 Strategies of observation, {\it e.g.} Fast-Switched
Phase referenced observations, or real-time measurements of these
fluctuations are being tested.  
This astrometric potential has the power of measuring orbital
parameters and masses of unseen companions as small as  0.1 Jupiter
for the closest stars with an observation program of 10 years.
The ALMA array  should start  operation
in 2006 with  8 antennas and gradually
increases to 64 antenna by 2010.        

ALMA will also be a powerful instrument to image protoplanetary disk, 
circumstellar debris disk and exozodiacal light (see Wootten in 
these Proceedings).

\begin{figure}[t]
\vspace{2.0cm}
\plotone{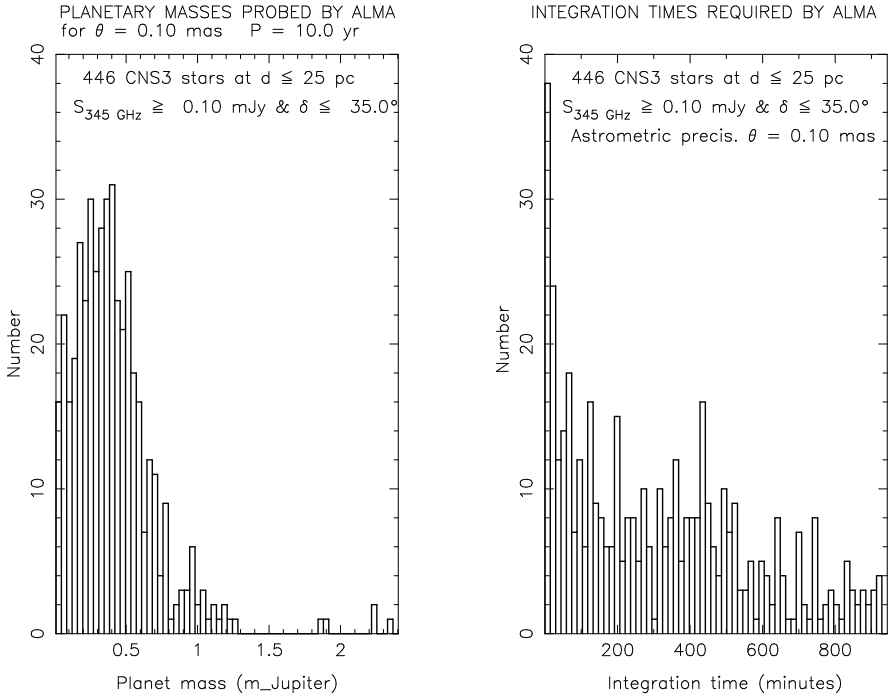}
\caption{\sl {\bf Left:} histogramme of minimum masses detectable astrometricly
by ALMA  for  possible unseen companions orbiting  nearby stars. 
The corresponding astrometric programme must last at least 10 years
 and maintains  an astrometric precision of 0.1~milliarcsecond. 
{\bf Right:} the integration times required to reach the theoretical 
astrometric precision of 0.1 milliarcsecond by ALMA.
}

\end{figure}

\null
\null

\noindent {\sl Acknowledgements :} I am grateful to Dr F. Crifo for 
clarifications on the spectral type nomenclature in the catalogue CNS3.

\null
\null

\noindent {\sl Bibliographical references : }

\noindent Gliese, W., Jarheiss, H., 1991,  http://www.ari.uni-heidelberg.de/aricns/files/cns3type.htm

\end{document}